\documentclass[reprint,amsmath,amssymb,aps,pra]{revtex4-2}
\usepackage{graphicx}
\usepackage{color}

\begin{document}

\title{Relationship between total reflection and Fabry-Perot bound states in the continuum}

\author{Zitao Mai}
\author{Ya Yan Lu}
\affiliation{Department of Mathematics, City University of Hong Kong, 
	Kowloon, Hong Kong, China} 
\date{\today}

\begin{abstract}
  Bound states in the continuum (BICs) have interesting properties and important applications in photonics. A particular class of BICs are found in Fabry-Perot (FP) cavities formed by two parallel periodic dielectric layers separated by a distance $h$. 
A periodic dielectric layer can totally reflect a plane incident wave with a particular frequency  and a particular wavenumber. Existing FP-BICs are found when $h$ is close to the values deduced from a phase-matching condition related to the reflection coefficient, but they are obtained in FP-cavities where the periodic layers have a reflection symmetry in the periodic direction. In this paper, we further clarify the connection between total reflections and FP-BICs. Our numerical results indicate that if the wavenumber is zero or the periodic layers have a reflection symmetry in the periodic direction, FP-BICs can indeed be found near the parameters of total reflections. However, if the wavenumber is nonzero and the periodic layer is asymmetric (in the periodic direction), we are unable to find a FP-BIC   (with a frequency and a wavenumber near those of a total reflection) by tuning $h$ or other structural parameters. Consequently, a total reflection does not always lead to a FP-BIC even when the parameters of the FP-cavity are tuned. 

\end{abstract}
\maketitle 

\section{Introduction}

In recent years, many studies on photonic bound states in the continuum (BICs) and their applications have been reported~\cite{mari08,bulg08,plot11,hsu13,kodi17,jin19}. In open linear wave systems, a BIC is  a localized eigenmode with a frequency in the radiation continuum~\cite{Hsu16}.
By perturbing the ideal structure with a BIC or changing the wavevector, resonant states  with arbitrarily high quality factors ($Q$ factors) can be created~\cite{yuan17pra,kosh18,yuan20}, leading to strong local field enhancement~\cite{moce15,hu20},  sharp features in transmission and reflection spectra~\cite{Blan16,Ship12,YZL22}, and
numerous applications in photonics~\cite{Sadr21,kosh23}. It is known that BICs are formed through different physical mechanisms, 
such as symmetry mismatch and destructive interference~\cite{Hsu16}. Some BICs are robust against small structural perturbations that preserve the relevant symmetry~\cite{zhen14,yuan17ol,yuan21}. A nonrobust BIC is usually destroyed by an arbitrary perturbation, but is may survive if  the perturbation contains one or more tunable parameters~\cite{Abdr22}. In structures with two periodic or translationally invariant directions, an electromagnetic BIC is a polarization singularity,  can be associated with an integer topological charge~\cite{zhen14,bulmak17}, and be observed as a polarization vortex in momentum space~\cite{doel18,zhang18}.
Some special BICs, the so-called super-BICs or merging BICs~\cite{yuan17pra,jin19,yuan20,kang21,Hwang21,kang22,bulg23}, 
turn to ultra-high-$Q$ resonances when the wavevector is perturbed, and they also exhibit bifurcations when the structure is perturbed~\cite{nan24}. 

A Fabry-Perot (FP) BIC is a special eigenmode  in a FP-cavity formed by two parallel thin mirrors separated by a distance, where the mirror can be a loseless dielectric periodic layer with one- or two-dimensional periodicity, such as a periodic array of dielectric cylinders~\cite{mari08,Ndan10} or a photonic crystal slab~\cite{Li16}. The BIC is trapped between two mirrors and has only an evanescent field outside the cavity. For the same frequency and wavevector as the BIC, plane waves can propagate to or from infinity and partially transmit through the mirrors.
It appears that FP-BICs are first found for linearized  water waves in two dimensions, where the ``mirrors'' are a pair of identical and infinitely long solid cylinders parallel to the water surface~\cite{mciver96,linton97,Porter02}. 
It is also known that FP-BICs exist in closed waveguides~\cite{chesnel}. 
In that case, the FP-cavity is formed by placing two scatterers in the waveguide core, and the propagating modes of the waveguide, instead of plane waves, serve as the radiation channels. The appearance of FP-BICs has been explained as the destructive interference of two resonances coupled to a single radiation channel~\cite{Hsu16}, but the theory is approximate and does not give definite predictions on the existence and non-existence  of FP-BICs. 

In between the two mirrors, the wave field of a FP-BIC is dominated by propagating  waves that are self-consistently  reflected by the mirrors. It is natural to look for FP-BICs around the frequency and wavevector at which a propagating wave is totally reflected by the mirror, and set the distance between the two mirrors by a phase matching condition related to the reflection coefficient~\cite{linton97,Porter02,mari08,chesnel}. However, due to the existence of evanescent waves, the FP-cavity with such a distance does not usually have a BIC.
When a plane wave impinges upon a periodic layer (i.e. the mirror), an infinite number of evanescent diffraction orders are excited. The evanescent waves excited at one mirror can generate transmitted and reflected propagating waves (of small amplitudes) at the other mirror, leading to radiations outside the FP-cavity. Therefore, the natural question is whether the FP-cavity with a slightly different distance can have a BIC. Existing studies indicate that the answer appears to be yes, but the results are either obtained numerically for special structures with particular parameters~\cite{linton97,Porter02,mari08,Ndan10} or asymptotically for special structures as the distance tends to infinity~\cite{chesnel}.
Moreover, the FP-BICs 
in closed waveguides are characterized by the frequency only, namely, they do not have a wavevector for the transverse direction.
symmetric structues.
The existing numerical studies for FP-BICs in periodic structures have focused on symmetric structures only~\cite{mari08,Ndan10,Li16}. 

In this paper, we present accurate  numerical results to further clarify the relation between total reflections and FP-BICs, taking into account the different symmetry of the structure, the wavevector for transverse spatial variables, and structural parameters in addition to the distance. In most cases, we show that FP-BICs indeed exist when the distance between the two mirrors or other parameters are slightly adjusted. But if the periodic structure does not have a reflection symmetry in the periodic direction and the wavenumber is nonzero, we show that FP-BICs may not exist when structural parameters are slightly tuned.

The rest of the paper is organized as followed. In Sec.~II, we recall the definitions and basic properties for diffraction and eigenvalue problems involving a periodic structure with a 1D periodicity, and clarify the assumptions about total reflections and BICs. In Sec.~III, we derive an approximate condition for FP-BICs using one propagating and one evanescent waves. Numerical examples for FP-BICs in periodic structures with different symmetry are presented in Sec.~IV.  The paper is concluded with some remarks in Sec.~V.

\section{Preliminaries}

\begin{figure}[ht]
  \centering
  \includegraphics[width=\linewidth]{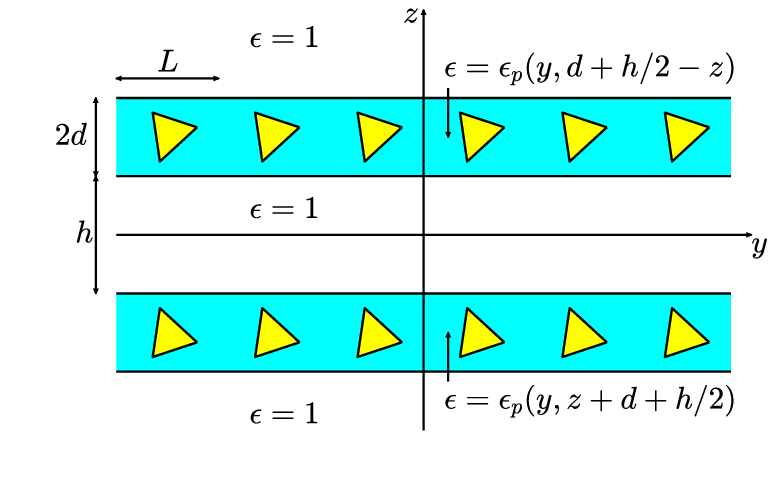}
  \caption{A Fabry-Perot cavity consisting of two periodic layers seperated by a distance $h$. The periodic layers are invariant in $x$, periodic in $y$ with period $L$, perpendicular to the $z$ axis, and have a thickness of $2d$. They are mirror reflections (with respect to the $z=0$ plane) of each other.}
  \label{FPstructure}
\end{figure}

In this section, we describe single-layer and double-layer (Fabry-Perot) structures  and recall the definitions and basic properties of total reflections and BICs. First, we define a dielectric periodic layer in a Cartesian coordinate system with  spatial variables $\{x,y,z\}$. The periodic layer is given for $-d < z < d$ by  a real dielectric function $\epsilon_{p}(y,z)$ which is independent of $x$ and periodic in $y$ with period $L$.  Assuming this periodic layer is surrounded by air, the dielectric function for the entire single-layer structure is then
\begin{equation}
  \label{1layer}
  \epsilon^{(1)}(y,z)=\begin{cases}
    \epsilon_p(y,z), & \quad    |z|< d,\\
    1, & \quad |z| >d.
  \end{cases}
\end{equation}
In Fig.~\ref{FPstructure}, we show a schematic of a FP-cavity.
It consists of two identical periodic layers separated by a distance $h$. These two layers are perpendicular to the $z$ axis with their middle planes at $z=\pm (h/2+d)$, respectively. In addition, we let the  upper periodic layer be the mirror reflection of the lower layer, so that the resulting FP-cavity has an up-down reflection symmetry. The dielectric function  of the entire double-layer structure is
\begin{equation}
  \epsilon^{(2)}(y,z)=\begin{cases}
    \epsilon_p(y, z + d + h/2), & -D < z< -h/2,\\
    \epsilon_p(y, d+h/2-z),  &  h/2 <z< D,\\
    1,  & \text{otherwise},
  \end{cases}
\end{equation}
where $D = 2d+h/2$.      

For both single- and double-layer structures, we consider time-harmonic  electromagnetic waves that are invariant in $x$ and depend on time as $e^{-i\omega t}$, where $\omega$ is the angular frequency. For the $E$-polarization, the $x$ component  of the electric field, denoted as $u$, satisfies the following Helmholtz  equation
\begin{equation}
  \label{helm}
  \partial_y^2u+\partial_z^2u+k_0^2\epsilon(y,z) u =0
\end{equation}
where $k_0=\omega/c$ is the free space wavenumber and $c$ is the speed of light in  vacuum, and $\epsilon$ is either $\epsilon^{(1)}$ or $\epsilon^{(2)}$.

For the single-layer structure, we consider the diffraction of a plane incident wave 
\begin{equation}
  \label{incident}
  u^{(i)} = e^{ i  [ \beta_0 y -  \gamma_0 (z-d) ] }, \quad z  > d, 
\end{equation}
specified above the layer, where $\beta_0 \in (-\pi/L, \pi/L]$,  $ |\beta_0| < k_0$, 
and $\gamma_0 = (k_0^2 - \beta_0^2)^{1/2} > 0$. The total wave $u$ satisfies Eq.~(\ref{helm}) for $\epsilon=\epsilon^{(1)}$, and can be written as
$u = u^{(i)} + u^{(r)}$ for $z > d$ and
$u = u^{(t)}$ for $z <- d$, respectively, where $u^{(r)}$ and $u^{(t)}$ are the reflected and transmitted waves, and they can be expanded as 
\begin{eqnarray}
  \label{refl0}
 u^{(r)} = \sum_{m= - \infty}^{\infty} R_{m,0} \, e^{i [ \beta_m y + \gamma_m (z-d)]} ,\quad z> d, \\
  \label{tran0}
 u^{(t)} = \sum_{m = -\infty}^{\infty} T_{m,0} \, e^{i [ \beta_m y - \gamma_m (z+d))] },\quad z <- d, \end{eqnarray}
where
\begin{equation}
  \label{defbtgm}
  \beta_m=\beta_0 + \frac{2\pi m}{L},  \quad \gamma_m=\sqrt{k_0^2-\beta_m^2}  
\end{equation}
are wavenumbers of the $m$-th diffraction order, $R_{m,0}$ and $T_{m,0}$ are the corresponding reflection and transmission coefficients. In the following, we assume
\begin{equation}
  \label{ass1ch}
  |\beta_0| < k_0 = \frac{\omega}{c} < \frac{2\pi}{L} - |\beta_0|.
\end{equation}
In that case, for all $m \ne 0$, $\gamma_m$ is purely  imaginary with a positive imaginary part,  the zeroth diffraction order is the only propagating order, and the power balance law implies that $|T_{0,0}|^2+|R_{0,0}|^2=1$.

It is known that $T_{0,0}$ can sometimes be exactly zero for a real frequency $\omega^*$ and a real wavenumber $\beta^*_0$~\cite{Popov86}.  In that case, $|R_{0,0}|=1$, and the power of the incident wave is completely converted to the reflected wave. This phenomenon, referred to as zero transmission or total reflection, is robust when the periodic layer has the relevant symmetry~\cite{Popov86,Blan16}. More precisely, if $\epsilon_p$ is symmetric in $y$ or $z$, namely, $\epsilon_p(-y,z) = \epsilon_p(y,z)$ or $\epsilon_p(y,-z)=\epsilon_p(y,z)$, and if zero transmission occurs at a real pair $(\omega^*, \beta^*_0)$, then for any sufficiently small loseless perturbation of $\epsilon_p$ that preserves the symmetry in $y$ or $z$, the zero transmission phenomenon still occurs, either at a frequency near $\omega^*$ for the same $\beta_0^*$, or  at a wavenumber near $\beta_0^*$ for the same $\omega^*$. It should be noted that robustness is not and does not imply existence. It is rather difficult to specify conditions on $\epsilon_p$ to guarantee the occurrence  zero-transmission phenomenon, but some limited results  are available~\cite{Ship12,chesnel,YZL22}. On the other hand, assuming the existence of a zero transmission, it is possible to derive approximate formulas for  $\omega^*$ or $\beta_0^*$~\cite{Fan03,Wu22}.

For the FP-cavity, we are interested in guided modes satisfying Eq.~(\ref{helm}) with $\epsilon=\epsilon^{(2)}$ and the boundary condition $u \to 0$ as $z \to \pm \infty$. Due to the periodicity in $y$, any guided mode is given in the Bloch form
\begin{equation}
  \label{Bloch}
  u(y,z) = e^{i \beta_0 y} \phi(y,z), 
\end{equation}
where $\beta_0 \in (-\pi/L, \pi/L]$ and $\phi$ is periodic in $y$ with period $L$. For $|z| > D$, we can expand $u$ as
\begin{equation}
  \label{Raymod}
  u(y,z) = \sum_{m=-\infty}^\infty c_m^{\pm} e^{ i ( \beta_m y \pm  \gamma_m z)}, \quad \pm z > D,
\end{equation}
where $\beta_m$ and $\gamma_m$ are given in Eq.~(\ref{defbtgm}).
Regular guided modes exist below the light line, i.e., $k_0 < |\beta_0|$, and they depend on $\beta_0$ and $\omega$ continuously. A BIC is also a guided mode, but it exists above the light line, i.e., $k_0 > |\beta_0|$. For simplicity, we assume condition (\ref{ass1ch}) is satisfied. In that case, $\gamma_0$ is positive and all $\gamma_m$ for $m\ne 0$ are purely imaginary. Since for any guided mode, $u \to 0$ as $z \to \pm \infty$, we must have  $c_0^{\pm} = 0$.

If $L$ is the true minimum period in $y$, i.e., the structure is not periodic with period $L/n$ for any integer $n\ge 2$, BICs  appear as isolated points in the $\beta_0$-$\omega$ plane. Let $\beta_0^\diamond$ and $\omega^\diamond$ be the Bloch wavenumber and the frequency of a BIC, then for a real $\beta_0$ near $\beta_0^\diamond$, Eq.~(\ref{helm}) with $\epsilon=\epsilon^{(2)}$ has a resonant state satisfying Eqs.~(\ref{Bloch}) and (\ref{Raymod}) with a complex  frequency $\omega$ and a nonzero $c_0^\pm$. Since the resonant state radiates power to $z =\pm \infty$ through the outgoing plane waves $e^{i (\beta_0 y \pm  \gamma_0 z)}$, it must decay with time and  has a negative $\mbox{Im}(\omega)$. Consequently, $\mbox{Im}(\gamma_0)$ is also negative and $e^{ \pm i \gamma_0 z}$ is unbounded as $z \to \pm \infty$.  The $Q$ factor of the resonant state, $Q = -0.5 \mbox{Re}(\omega)/\mbox{Im}(\omega)$, usually satisfies
$Q \sim 1/|\beta_0 - \beta_0^\diamond|^2$, but for some special BICs (the so-called super-BICs), $Q \sim 1/|\beta_0-\beta_0^\diamond|^m$ for $m\ge 4$.

If the single periodic layer exhibits a total reflection with frequency $\omega^*$ and wavenumber $\beta_0^*$, we can determine the distance $h$ of the FP-cavity, such that the plane wave between the mirrors is self-consistently reflected. Let the reflection coefficient be $R_{0,0} = e^{ i \theta^*}$ for some $\theta^* \in (-\pi, \pi]$ and $\gamma_0^* = \sqrt{ (\omega^*/c)^2 - (\beta_0^*)^2}$. When the plane wave is reflected by the lower mirror at $z=-h/2$, travels to the upper mirror, is reflected by the upper mirror at $z=h/2$, and travels down to the lower mirror again, it acquires the phase  $2(\theta^* +  \gamma_0^* h)$ which must be $2\pi n$ for some integer $n$. Therefore, the plane wave is consistently reflected if the distance $h$ satisfies 
\begin{equation}
  \label{formulah}
  h = h_n^* := \frac{ n \pi - \theta^*}{\gamma_0^*}, 
\end{equation}
where $n$ is any integer such that $h_n^*$ is positive. However, this does not imply that the FP-cavity with $h=h_n^*$ has a BIC with frequency $\omega^*$ and Bloch wavenumber $\beta_0^*$. Evanescent waves that decay towards a periodic layer can actually generate propagating waves. Thus, the total propagating plane wave with frequency $\omega^*$ and wavenumber $\beta_0^*$ is no longer totally reflected by the mirrors. In fact, the FP-cavity with $h=h_n^*$ usually does not even have a BIC with wavenumber $\beta_0=\beta_0^*$ and  frequency $\omega$ near $\omega^*$. In the following, we consider the distance $h$ and other structural parameters, and investigate  whether the FP-cavity has a BIC with $\omega$ near $\omega^*$ and $\beta_0$ near or equal to $\beta_0^*$ when $h$ or other parameters are slightly varied.

\section{Approximation with one evanescent wave} 

Before we present numerical results for FP-BICs with distance, frequency and wavenumber near or at $h_n^*$, $\omega^*$ and $\beta_0^*$ associated with a total reflection, we first consider a model assuming that the wave field in the FP-cavity consists of one propagating and one evanescent waves. The model helps us to understand why the FP-BIC, if it exists, does not have the exact total-reflection values $h_n^*$, $\omega^*$ and $\beta_0^*$. It also gives rise to approximate formulas for $h$ and $\omega$ (or $\beta_0$) of the FP-BIC. 

For the single periodic layer bounded by $z=\pm d$, if $\beta_0 \in (0, \pi/L)$, the negative first diffraction order has the slowest decay rate in $z$. If we replace the incident wave in Eq.~(\ref{incident}) by an evanescent wave of diffraction order $-1$, i.e.
\begin{equation}
  \label{incident1}
  u^{(i)} = e^{  i  [ \beta_{-1} y -  \gamma_{-1} (z-d) ] }, \quad z  > d, 
\end{equation}
then the reflected and transmitted waves are
\begin{eqnarray}
  \label{refl1}
   u^{(r)} = \sum_{m= - \infty}^{\infty} R_{m,-1} \, e^{i [ \beta_m y + \gamma_m (z-d)]} ,\quad z> d, \\
  \label{tran1}
   u^{(t)} = \sum_{m = -\infty}^{\infty} T_{m,-1} \, e^{i [ \beta_m y - \gamma_m (z+d))] },\quad z <- d, 
\end{eqnarray}
where $T_{m,-1}$ and $R_{m,-1}$ are transmission and reflection coefficients for the $m$-th diffraction order. 

Now, we assume the FP-cavity  has a BIC and its wave field between the two mirrors is 
\begin{eqnarray}
  \label{onep1}
  \nonumber
&&  u  \approx 
a_0^+\ e^{i [ \beta_0 y+\gamma_0 (z-h/2)] } +a_{-1}^+\
  e^{i [ \beta_{-1} y+\gamma_{-1} (z-h/2)]} \\
&& +a_0^- \ e^{i [ \beta_0 y+\gamma_0 (z+ h/2)] } +a_{-1}^-\
  e^{i [ \beta_{-1} y+\gamma_{-1} (z+h/2)]} 
\end{eqnarray}
for $-h/2 < z < h/2$.
The total transmitted propagating wave for $z>D$ is
$
  (  a_0^+\, T_{0,0} + a_{-1}^+\, T_{0,-1}  ) e^{ i [ \beta_0 y + \gamma_0 (z-D)]}.
$
Thus an approximate condition for the FP-BIC is
\begin{equation}
  \label{apc1}
   a_0^+\, T_{0,0} + a_{-1}^+\, T_{0,-1}  =0.
 \end{equation}
 However, the above condition is not convenient to use, since $a_0^+$ and  $a_{-1}^+$ are still undetermined.

 We can find an alternative condition based on the wave field in the cavity being self-consistently reflected. 
At $z=h/2$, the amplitudes of the reflected zeroth and negative first diffraction orders are related to the amplitudes of the incident waves by 
\begin{equation}
{\bf D}^{-1} 
\left[ \begin{array}{c}
          a_0^- \\
          a_{-1}^- 
       \end{array}
     \right]
     =
     {\bf R} 	\left[\begin{array}{c}
		a_0^+ \\
		a_{-1}^+
              \end{array}\right], 
\end{equation}
where
\begin{equation}
  {\bf D}  = 
  \left[\begin{array}{cc}
          e^{ i \gamma_0 h} & 0\\
          0 & e^{i \gamma_{-1} h}
        \end{array}\right], \quad 
  {\bf R}  = 
  \left[\begin{array}{cc}
          R_{0,0}&R_{0,-1}\\
          R_{-1,0}&R_{-1,-1}
        \end{array}\right].
\end{equation}    
Similarly, at $h=-h/2$, we have 
\begin{equation}
{\bf D}^{-1} 
\left[ \begin{array}{c}
          a_0^+ \\
          a_{-1}^+
       \end{array}
     \right]
     =
     {\bf R} 	\left[\begin{array}{c}
		a_0^- \\
		a_{-1}^-
              \end{array}\right].
\end{equation}
Actually, the FP-BIC is either even or odd in $z$, thus
$( a_0^-, a_{-1}^- ) = \pm ( a_0^+, a_{-1}^+ )$. Therefore, either
${ \bf D} {\bf R} -{\bf I}$ or $ {\bf D} {\bf R} + {\bf I}$ is a singular matrix, where
${\bf I}$ is the $2\times 2$ identity matrix. We can write the  approximate condition for  FP-BICs as
\begin{equation}
  \label{drpm1}
\det ( {\bf D}{\bf R} \pm {\bf I}) = 0
\end{equation}
or $\det [ ({\bf D} {\bf R})^2 - {\bf I} ] = 0$. If we keep only the propagating wave in the  cavity, then ${\bf D}$ and ${\bf R}$ become their $(1,1)$ entries, and the condition is simplified to 
$  [ e^{ i \gamma_0 h} R_{0,0} ]^2 = 1$
This immediately leads to Eq.~(\ref{formulah}).

Since Eq.~(\ref{drpm1}) is a complex equation, it can be used to calculate two real quantities, for example $\omega$ and $h$, assuming $\beta_0$ is fixed. Moreover, we can further simplify Eq. ~(\ref{drpm1}) to obtain approximate linear equations for $h$ and $\omega$. Let $(\omega^*, \beta_0^*)$ correspond to  a total reflection and $h^* = h_n^*$ be given by Eq.~(\ref{formulah}) for an even integer $n$, then the BIC 
(if it exists) should be even in $z$, so that $\det ( {\bf D} {\bf R} - {\bf I}) = 0$.
Taking the partial derivative of this equation with respect to $\omega$ and $h$ at $(\omega^*, h^*)$, we have
\begin{eqnarray}
\nonumber  && \left[ h^* \omega^*/( c^2 \gamma^*_0)  - R'_{00}(\omega^*) \right] 
              (\omega - \omega^*) \\
  \label{1storder}
&&  - \gamma_0^* (h - h^*)  \approx  i e^{ i \gamma_{-1}^* h^*} R_{0,-1}^* R_{-1,0}^* /  R_{00}^*
\end{eqnarray}
where $R_{j,k}^* = R_{j,k}(\omega^*)$, $R_{00}'(\omega^*)$ is the derivative of $R_{00}$ with respect to $\omega$  at $\omega^*$, $\gamma_j^* = [ (\omega^*/c)^2 - (\beta_j^*)^2 ]^{1/2}$ for $j, k \in \{ 0, -1\}$. The real and imaginary parts of Eq.~(\ref{1storder}) form a linear system of two equations for $h$ and $\omega$. If the BIC is odd in $z$, we can derive a similar linear system from $\det( {\bf D}{\bf R} + {\bf I})=0$.

\section{Numerical Examples}

\begin{figure}[h]
  \centering 
  \includegraphics[width=0.9\linewidth]{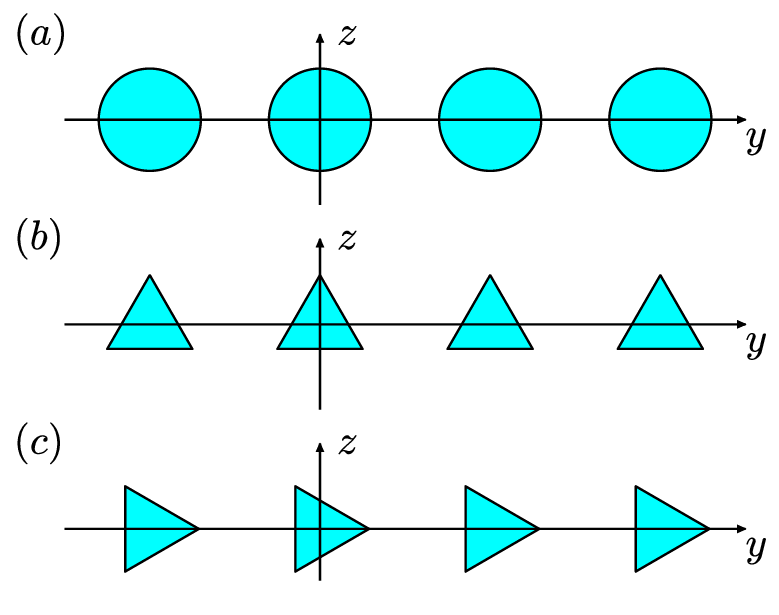}
  \caption{Three periodic arrays of cylinders with period $L$ in the $y$ direction. Cross sections of the cylinders: (a) circular disks with radius $a$,   (b) equilateral triangles with side length $L_t$ and a reflection symmetry in $y$, (c) equilateral triangles with side length $L_t$ and a reflection symmetry in $z$.}
  \label{fig:Model}
\end{figure}

In this section, we present numerical examples for BICs in FP-cavities constructed from two periodic arrays of circular or triangular cylinders separated by a distance $h$. In Fig.~\ref{fig:Model},
we show a periodic array of circular cylinders with radius $a$ and two periodic arrays of 
triangular cylinders that are symmetric in $y$ and $z$, respectively. The cross sections of  all triangular cylinders are equilateral triangles of side length $L_t$. The dielectric constant of all circular or triangular cylinders is $\epsilon_{c} = 6.6$. The medium outside the cylinders is air. All three periodic arrays of cylinders are regarded as a periodic layer of thickness $2d=L$, where $L$ is the period of the array in the $y$  direction.

For each periodic array, we first calculate two total reflections with a zero and a nonzero  $\beta_0^*$, respectively, then for each total reflection, we select a distance $h_n^*$, and try to find FP-BICs by varying the Bloch wavenumber $\beta_0$, the distance $h$ or other structural parameters such as $a$ and $L_t$. The numerical method is based on a nonlinear eigenvalue formulation at $z=\pm D$ for one period of $y$~\cite{hu15,Yuan17} and  a contour integral method for solving nonlinear eigenvalue problems~\cite{asak09,beyn12}.

\subsection{Circular cylinders}

For the periodic array of circular cylinders with $a=0.3L$ and shown in Fig.~\ref{fig:Model}(a), we found two total reflections with $\beta_0^* =0$ and $\beta_0^* = 0.5 (\pi/L)$,  respectively, and they are listed in Table~\ref{circular1}
\begin{table}[h]
  \centering 
  \begin{tabular}{c||c|c|c|c} \hline 
   Solutions &\ $a/L$\  & $h/L$ & $\beta_0 L/\pi$ & $\omega L/(2\pi c)$ \\ \hline\hline 
    TR1 & 0.3 & \ 0.43394 \  & 0 & 0.99310 \\ \hline 
    RM1 & 0.3 & 0.43394 & 0 & 0.99330$-$0.000039i \\ \hline 
    BIC1 & 0.3 & {\sf 0.41637} & 0 & 0.99557 \\ \hline 
    BIC2 & \ {\sf 0.30152}\  & 0.43394 & 0 & 0.99069 \\ \hline\hline 
    TR2 & 0.3 & \ 0.62438 \  & 0.5  & 0.70413 \\ \hline 
    RM2 & 0.3 & 0.62438 & 0.5 & 0.70679$-$0.000072i \\ \hline 
    BIC3 & 0.3 & {\sf 0.58363} & 0.5 & 0.71054 \\ \hline 
    BIC4 & \ {\sf 0.30816}\  & 0.62438 & 0.5 & 0.69511 \\ \hline 
    BIC5 & 0.3 & 0.62438 & {\sf 0.51884}\  & 0.70511 \\ \hline 
  \end{tabular}
  \caption{FP-cavity with two periodic arrays of circular cylinders of radius  $a$ separated by a distance $h$. The different rows show total reflections, resonant modes and BICs.}
  \label{circular1}
\end{table}
as TR1 and TR2. For each total reflection, we choose a distance $h_n^*$, denoted as $h^*$ for simplicity, from formula (\ref{formulah}) for a small integer $n$, to form a Fabry-Perot cavity. The cavity with distance $h^*$ has only a resonant mode with a complex frequency and Bloch wavenumber $\beta_0 = \beta_0^*$. The two resonant modes are listed in
Table~\ref{circular1} as RM1 and RM2, respectively.

\begin{figure}[h]
  \centering
  \includegraphics[width=\linewidth]{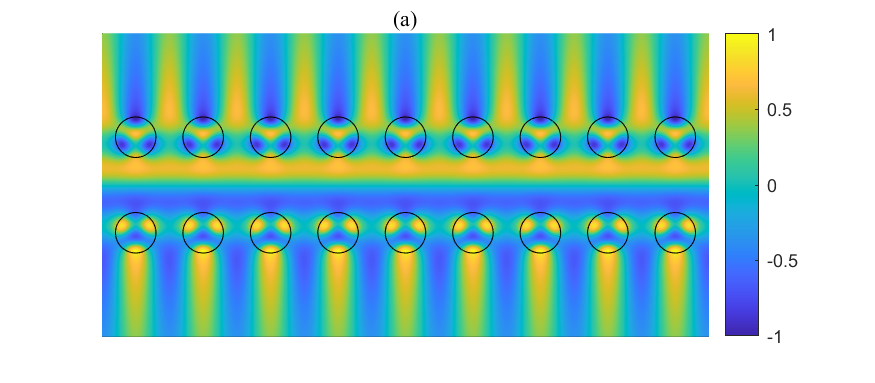}\\
  \includegraphics[width=\linewidth]{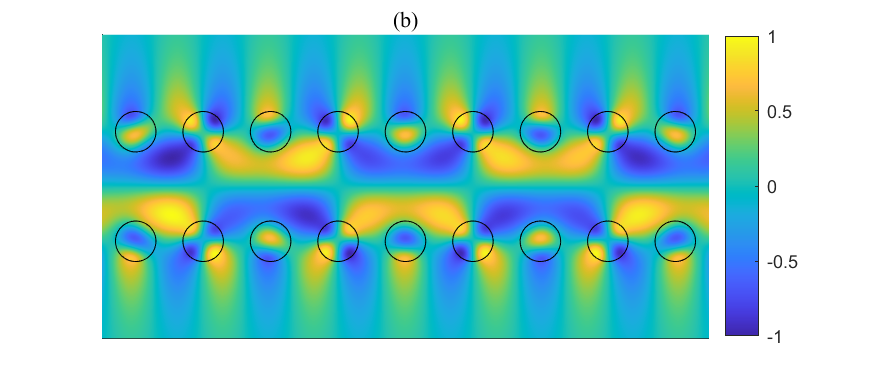}  
  \caption{Field patterns (real part of $u$) of BIC1 and BIC5 as in Table~\ref{circular1}.}
  \label{fieldA}
\end{figure}

To find BICs in this Fabry-Perot structure, we have to change a structural parameter or vary the Bloch wavenumber $\beta_0$. For the total reflection with $\beta_0^* = 0$, we found two FP-BICs, BIC1 and BIC2 in Table~\ref{circular1}, by changing the distance  from $h^*=0.43394L$ to $h=0.41637L$ and changing the radius $a$ from $0.3L$ to
$0.30152L$, respectively. The frequencies of these two BICs are given in Table~\ref{circular1}. For the total reflection with $\beta_0^* = 0.5 (\pi/L)$, we similarly found two FP-BICs, listed in Table~\ref{circular1} as BIC3 and BIC4, by changing $h$ and $a$, respectively. In that case, it is also possible to find a FP-BIC by keeping $h=h^*$ and $a=0.3L$ and changing $\beta_0$. The result is the
BIC5 given in Table~\ref{circular1}. The wave field patterns of BIC1 and BIC5 are shown in Fig.~\ref{fieldA}.

The FP-BICs listed in Table~\ref{circular1} are consistent with existing theories about their robustness and parametric dependence~\cite{yuan17ol,Abdr22}.  Since the FP-cavity has reflection symmetry in both $y$ and $z$, a typical BIC with its frequency $\omega$ and a nonzero Bloch wavenumber $\beta_0$ satisfying (\ref{ass1ch}) is robust. This means that when a structural parameter is slightly perturbed without breaking the symmetry, the BIC will continue to exist, but will have a slightly different frequency and a slightly different wavenumber~\cite{yuan17ol}. For example, when $h$ is decreased from $0.62438L$ to $0.58363L$, BIC5 is turned to BIC3, and when $a$ is increased from $0.3L$ to $0.30816L$, BIC5 is turned to BIC4. The robustness theory also implies that if 
$\beta_0$ is fixed and  one structural parameter is changed, the BIC can be preserved if another structural parameter is properly adjusted. For example, if  $a$ is slightly increased and $h$ is changed accordingly, BIC3 is turned to BIC4 with $\beta_0$  fixed at $0.5 (\pi/L)$.

The two FP-BICs with $\beta_0=0$, i.e., BIC1 and BIC2 in Table~\ref{circular1}, are not typical BICs. For structures with a reflection symmetry in $y$, the typical BICs with $\beta_0=0$ are symmetry-protected standing waves that are anti-symmetric in $y$. Both BIC1 and BIC2 are symmetric standing waves (symmetric in $y$),  unprotected by symmetry, and nonrobust~\cite{yuan17pra}. In fact, they are the simplest super-BICs with the $Q$ factor of nearby resonant modes proportional to $1/\beta_0^4$. Nevertheless, by adjusting a second parameter, such a BIC can still exist continuously as one structural parameter is varied. For example, when $a$ is increased from $0.3L$ to $0.30152L$ and $h$ is increased from $0.41637L$ to $0.43394L$, BIC1 is turned to BIC2, and both are standing waves with $\beta_0=0$.

\subsection{Triangular cylinders: symmetric in $y$}

\begin{figure}[h]
  \centering 
  \includegraphics[width=\linewidth]{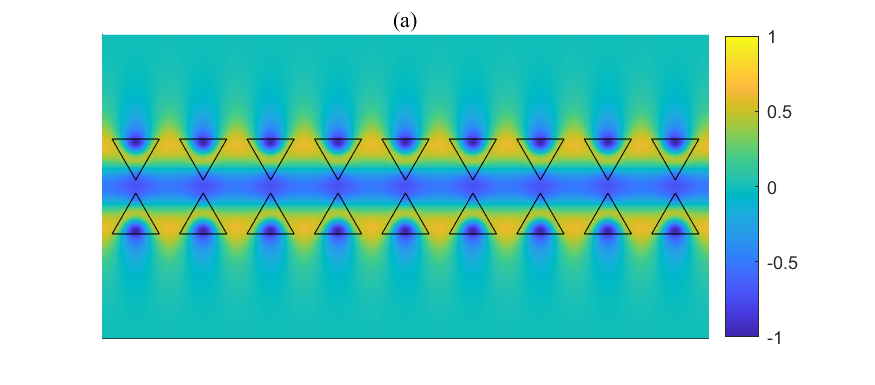} \\
  \includegraphics[width=\linewidth]{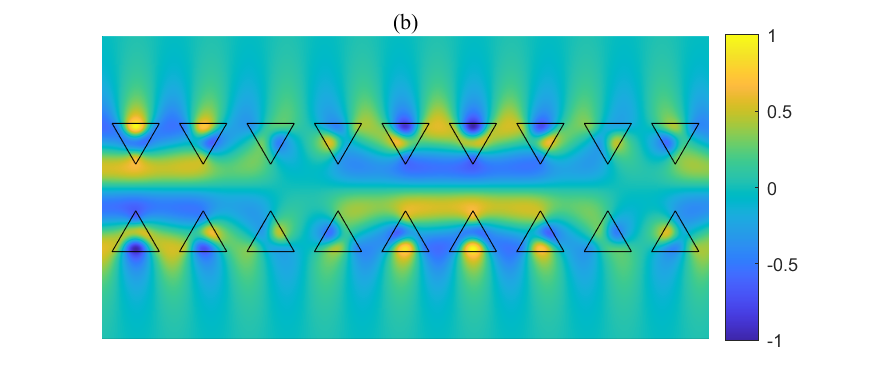}
  \caption{Field patterns (real part of $u$) of BIC2 and BIC4 as in Table~\ref{triang1}.}
  \label{fieldB}
\end{figure}

For the periodic array of triangular cylinders 
shown in Fig.~\ref{fig:Model}(b), we assume the side length is
$L_t = 0.7L$ and found two total reflections with $\beta_0^*=0$ and $\beta_0^* = 0.2 (\pi/L)$, respectively. For each case, the FP-cavity with a distance $h^*$, chosen from formula (\ref{formulah}) for a small integer $n$, does not have a BIC with the same Bloch wavenumber $\beta_0^*$. Instead, the cavity only has resonant modes with a complex frequency. In Table~\ref{triang1},
\begin{table}[h]
  \centering 
  \begin{tabular}{c||c|c|c|c} \hline 
   Solutions &\ $L_t/L$\  & $h/L$ & $\beta_0 L/\pi$ & $\omega L/(2\pi c)$ \\ \hline\hline 
    TR1 & 0.7 & \ 0.00272 \  & 0 & 0.83010 \\ \hline 
    RM1 & 0.7 & 0.00272  & 0 & 0.83007$-$0.000097i \\ \hline 
    BIC1 & 0.7 & \ {\sf 0.02274}\  & 0 & 0.82634 \\ \hline 
    BIC2 & \ {\sf 0.69045}\  & 0.00272 & 0 & 0.83774 \\ \hline 
    BIC3 & 0.7 & 0.00272 & $\pm${\sf 0.10028}  & 0.82958  \\ \hline \hline 
    TR2 & 0.7 & \ 0.48667 \  & 0.2  & 0.84515 \\ \hline 
    RM2 & 0.7 & 0.48667 & 0.2 & 0.84546$-$0.000011i \\ \hline 
    BIC4 & 0.7 & {\sf 0.50135} & 0.2 & 0.84354 \\ \hline 
    BIC5 & \ {\sf 0.69630}\  & 0.48667 & 0.2 & 0.84830\\ \hline 
    BIC6 & 0.7 & 0.48667 & {\sf 0.20895}\  & 0.84495 \\ \hline 
  \end{tabular}
  \caption{FP-cavity with two periodic arrays of triangular cylinders with side length $L_t$ and a reflection symmetry in $y$, separated by a distance $h$. The different rows show total reflections, resonant modes, and BICs.}
  \label{triang1}
\end{table}
the rows with RM1 and RM2 show $L_t$, $h^*$, $\beta_0^*$ and the complex frequency of the corresponding resonant modes. As in the previous subsection, 
we can find BICs near the total reflections by changing $h$, changing $L_t$ or varying $\beta_0$. The results are listed in Table~\ref{triang1}.
The wave field patterns of BIC2 and BIC4 are shown in Fig.~\ref{fieldB}.

While the single periodic array of triangular cylinders considered in this subsection does not have a reflection symmetry in $z$, the FP-cavity, consisting  of two arrays that are mirror reflection of each other, has reflection symmetry in both $y$ and $z$. As such, we have the same statements about BIC4, BIC5 and BIC6 as those for BIC3, BIC4 and BIC5 of Table~\ref{circular1}, and the same statements about BIC1 and BIC2 as in the previous subsection. In particular, the symmetric standing waves BIC1 and BIC2 are not typical BICs. For a structure with such a BIC, if a parameter is slightly changed,  the perturbed structure does not usually have a symmetric standing wave, and may or may not have a pair of BICs with nonzero and opposite $\beta_0$.  For example, when $h$ is changed from $0.02274L$ to $0.00272L$, the symmetric standing wave BIC1 is destroyed, but the perturbed system has a pair of BICs with nonzero and opposite $\beta_0$, listed in Table~\ref{triang1} as BIC3. On the other hand, for the FP-cavity with BIC1 of Table~\ref{circular1} in subsection A, if the distance is increased from $h=0.41637L$ to $0.43394L$, the symmetric standing wave is destroyed, and there are no other BICs with small but nonzero $\beta_0$. The existence or nonexistence of BICs with nonzero $\beta_0$ can be explained by the bifurcation theory for such non-generic BICs~\cite{nan24}.

\subsection{Triangular cylinders: asymmetric in $y$}

In this subsection, we consider the FP-cavity formed by two
periodic arrays of triangular cylinders shown in Fig.~\ref{fig:Model}(c).
The periodic arrays and the FP-cavity have a reflection
symmetry in $z$, but are asymmetric in $y$. For the single periodic
array with $L_t=0.5L$, we found two total reflections with $\beta_0^* = 0$ and $\beta_0^*=  0.2 (\pi/L)$, respectively. For each case, we choose a distance $h_n^*$ for a small integer $n$, denoted as $h^*$ for simplicity, and calculate the associated resonant mode.  The total reflections and the 
resonant modes are listed in Table~\ref{triang2}
\begin{table}[h]
  \centering 
  \begin{tabular}{c||c|c|c|c} \hline 
   Solutions &\ $L_t/L$\  & $h/L$ & $\beta_0 L/\pi$ & $\omega L/(2\pi c)$ \\ \hline\hline 
    TR1 & 0.5 & \ 0.64101 \  & 0 & 0.74436\\ \hline 
    RM1 & 0.5 &   0.64101  & 0 &   0.74413$-$i0.61$\times$$10^{-6}$ \\ \hline 
    BIC1 & 0.5 & {\sf 0.64732}\  & 0 & 0.74409 \\ \hline 
    BIC2 & \ {\sf 0.49248}\  & 0.64101 & 0 & 0.75165 \\ \hline \hline
    TR2 & 0.5 & \ 0.11847 \  & 0.2  & 0.70935 \\ \hline 
    RM2 & 0.5 & 0.11847 & 0.2 & 0.71161$-$i0.89$\times$$10^{-4}$ \\ \hline 
  \end{tabular}
  \caption{FP-cavity with two periodic arrays of
    triangular cylinders with no reflection symmetry in $y$. The different rows show total reflections, resonant modes and BICs.}
  \label{triang2}
\end{table}
as TR1, TR2, RM1 and RM2, respectively.

\begin{figure}[h]
  \centering 
  \includegraphics[width=\linewidth]{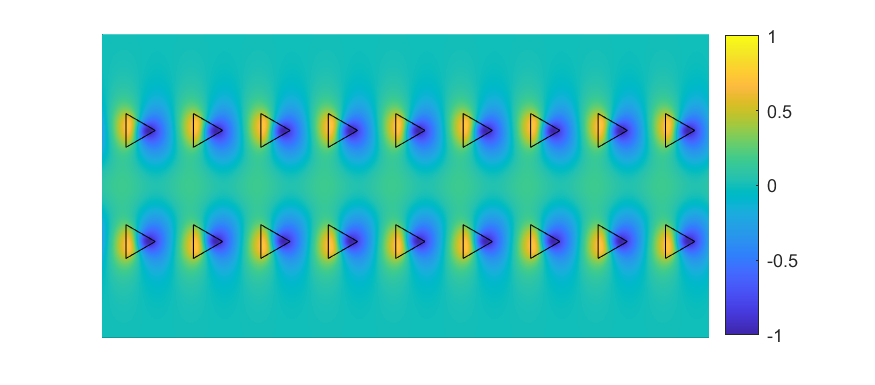}
  \caption{Field pattern (real part of $u$) of BIC1  in Table~\ref{triang2}.}
  \label{fieldC}
\end{figure}

In connection with the total reflection
at $\beta_0^* = 0$, we are able to find BICs by slightly changing $h$
or $L_t$. The results are listed in Table~\ref{triang2} as BIC1 and
BIC2.
The wave field pattern of BIC1 is shown in Fig.~\ref{fieldC}. 
Notice that all BICs with $\beta_0=0$ in periodic structures
without a reflection symmetry in $y$ are nonrobust. If a structural
parameter is varied, such a BIC will normally be destroyed. But if a second parameter is adjusted accordingly following the first one, the BIC can exist continuously. For
example, as $L_t$ is reduced from $0.5L$ to $0.49248L$ and $h$ is also
changed from $0.64732L$ to $0.64101L$, BIC1 is turned to BIC2.

\begin{figure}[h]
	\centering 
\includegraphics[width=\linewidth]{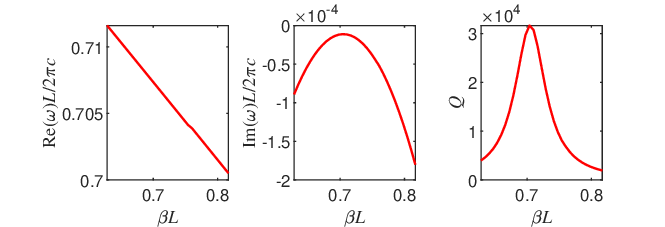}
\caption{Complex frequency and $Q$ factor of the resonant modes in an asymmetric FP-cavity with
  distance $h^*=0.11857L$,  for $\beta_0 \ge 0.2\pi/L$.}
\label{nbicbeta}
\end{figure}

For the total reflection at $\beta_0^* = 0.2 (\pi/L)$, the FP-cavity with the distance $h^* = 0.11847L$ does not have a BIC for any $\beta_0$ near $\beta_0^*$. In Fig.~\ref{nbicbeta},
we show the real and imaginary parts of the complex frequency $\omega$ and the $Q$ factor of the corresponding resonant modes  for $ \beta_0  L/\pi \in [0.2, 0.26]$. It is clear that
$\mbox{Im}(\omega)$ has a local maximum which is negative, and the $Q$ factor is always finite. According to the theory
on parametric dependence of BICs~\cite{Abdr22}, a BIC in such a periodic structure without a
reflection symmetry in $y$ can form a curve in the plane of two generic structural 
parameters.  This implies that in
the $h$-$L_t$ plane, there maybe a curve, and at each point on the curve the FP-cavity has a BIC for some $\omega$ and some $\beta_0$. Since the probability of a random point falling on a curve is zero, it is not surprising that no BIC can be found for $L_t=0.5L$ and $h=h^*$. 

\begin{figure}[h]
	\centering 
\includegraphics[width=\linewidth]{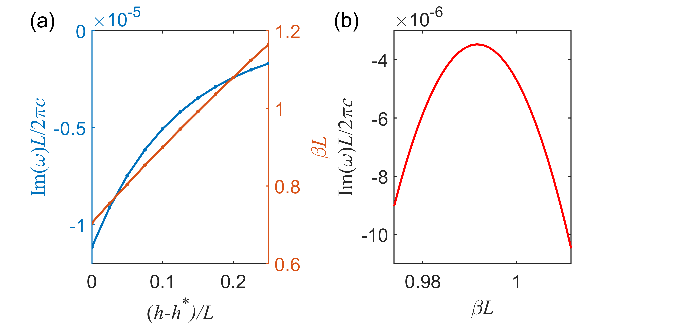}
\caption{(a) The local maximum of the imaginary part of the frequency of the resonant mode and the related wavenumber $\beta$ coming from adding a $d$ on $h$. (b) When $d=0.03$, the trajectory of the imaginary part of the frequency of the resonant mode for varying $\beta$.}
	\label{nbich}
\end{figure}

\begin{figure}[h]
	\centering 
\includegraphics[width=\linewidth]{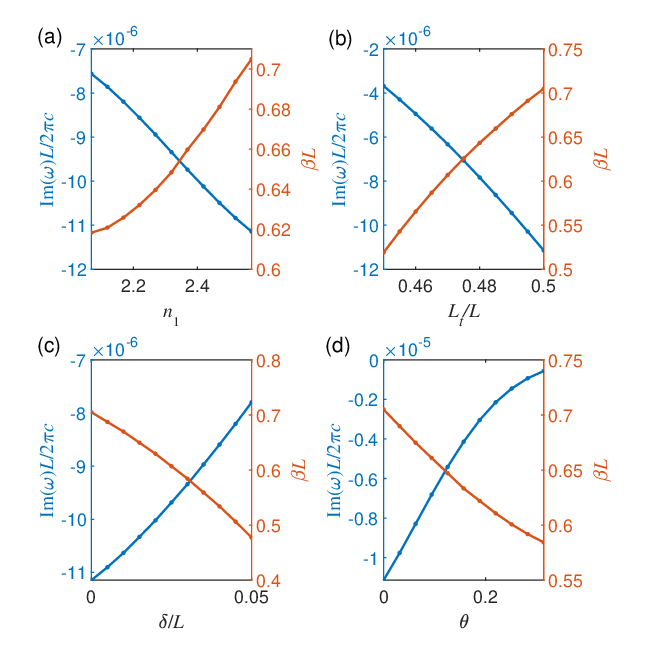} 
\caption{Local maximum of $\mbox{Im}(\omega)$ and $\beta_0$ at the local maximum for resonant modes in a FP-cavity with a varying (a) refractive index $n_1$, (b)  side length $L_t$, (c) height $H_t$ or $\delta = H_t - (\sqrt{3} /2) L_t$,  and  (d) rotation angle $\theta$.}
\label{nbic4}
\end{figure}

To find out whether BICs exist near this total reflection with nonzero $\beta_0^*$ when the structural is slightly changed, we consider
five parameters: distance $h$, side length $L_t$, refractive index of the
cylinders 
$n_1 = \sqrt{\epsilon_c}$, height $H_t$ of the triangle (cross section of the
cylinders) in the $y$ direction, and rotation angle $\theta$ of the triangle about its center, where $H_t$ and $\theta$ are
\begin{figure}[th]
  \centering
\includegraphics[width=\linewidth]{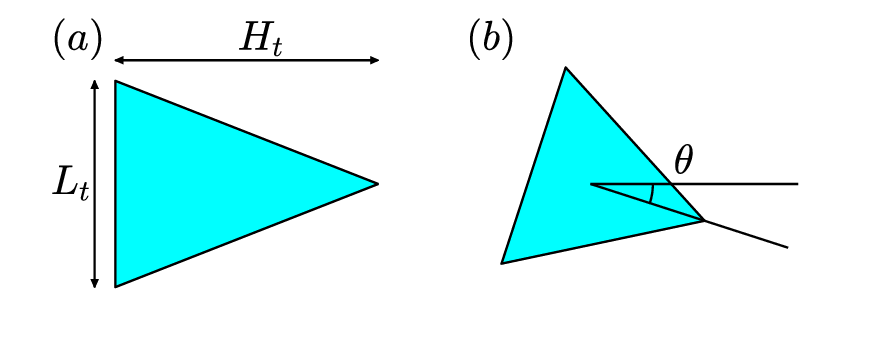}
  \caption{(a) An isosceles triangle with base $L_t$ and height $H_t$. (b) An equilateral triangle rotated by angle $\theta$.}
  \label{2triangles}
\end{figure}
 illustrated in Fig.~\ref{2triangles}.
When each of these parameters is varied, we search the 
wavenumber $\beta_0$ around $\beta_0^*$ for a local maximum of
$\mbox{Im}(\omega)$, where $\omega$ is the complex frequency of the 
resonant modes. A BIC can be identified if $\mbox{Im}(\omega) =
0$. However, in all five cases, we are unable to find a BIC with $\omega$ near $\omega^*$ and $\beta_0$ near $\beta_0^*$. 
In Fig.~\ref{nbich},
we show the result for varying $h$ from $h^*$ to $h^* + 0.25L$.
A particular case for $h= h^* + 0.015L$ is shown in Fig.~\ref{nbich}(b). 
Similar to the case for $h=h^*$ shown in the middle panel of Fig.~\ref{nbicbeta}, $\mbox{Im}(\omega)$ has a local maximum, which is still negative, at some $\beta_0 > \beta_0^*$. The maximum value of
$\mbox{Im}(\omega)$ and the $\beta_0$ that attains the local maximum are shown in
Fig.~\ref{nbich}(a) as functions of $h-h^*$. 
We only show the results for $h > h^*$, since the local maximum of $\mbox{Im}(\omega)$ increases as $h$ is increased from $h^*$. Since the studied ranges of $h$ and $\beta_0$ are quite large compared with $h^*$ and $\beta_0^*$, we conclude that this FP-cavity has no BIC for $h$ near $h^*$ and $\beta_0$ near $\beta_0^*$. Similar results are shown in 
Fig.~\ref{nbic4},
where each of these four parameters, $n_1$, $L_t$, $H_t$ and $\theta$, is varied. 
%
%
%
For all cases, we are unable to find a BIC even when one structural
parameter has varied a large amount.

It appears that for FP-cavities without a reflection symmetry in $y$, it is unlikely (at least not always possible) to find a
BIC with $\omega$ near $\omega^*$ and  $\beta_0$ near $\beta_0^*$ if $\beta^*_0$ is nonzero, by varying one structural parameter. As we  mentioned
earlier, if the structure has a BIC, then the BIC should form a curve
in the plane of two generic structural parameters~\cite{Abdr22}. If an initial pair of
parameters related to a total reflection is far away
from the curve, we will not be able to reach the curve by a small
variation of the parameters. Even if the initial pair is close to the
curve, and we are able to move to the curve by varying one parameter,
the $\beta_0$ of the BIC on the curve could be very different from $\beta_0^*$ of
the total reflection. In that case, there is still no BIC with $(\omega, \beta_0)$ near $(\omega^*, \beta_0^*)$.

\section{Conclusion}

In this work,  we  investigated the existence of FP-BICs near
total reflections
in Fabry-Perot cavities formed by two periodic arrays of circular or triangular dielectric cylinders separated by a distance $h$. A total
reflection of a single periodic array is associated with a frequency $\omega^*$ and a wavenumber
$\beta_0^*$. Our numerical results
indicate that the reflection symmetry in $y$ plays a key role and
whether $\beta_0^*$ is zero or nonzero is also important. More
specifically, if the periodic array is symmetric in $y$, it is likely 
to find a BIC with frequency $\omega$ near $\omega^*$ and Bloch
wavenumber $\beta_0 = \beta_0^*$  by slightly tuning one structural
parameter, such as the distance $h$, that does not break the symmetry in $y$.
For the case of $\beta_0^* \ne 0$, it is also likely to find a
BIC with $\beta_0$ near $\beta_0^*$ without tuning any structural
parameter. If the array does not have a reflection symmetry in $y$ and
$\beta_0^* = 0$, it is still likely to find a BIC with $\beta_0=0$ by tuning one structural
parameter. But if the array is not symmetric in $y$ and 
$\beta_0^* \ne 0$, it is unlikely to find a BIC with $(\omega,
\beta_0)$ near $(\omega^*, \beta_0^*)$ by varying any structural
parameter. The last case indicates that a total reflection does not
always lead to a FP-BIC, even when the parameters of the FP-cavity are slightly tuned. 

The FP-BICs found in Sec.~IV, listed in Tables I, II and III, belong to
different categories. They can be robust or nonrobust, and the nonrobust BICs include ordinary ones and super-BICs.
Those BICs with a nonzero $\beta_0$ in structures
with a reflection symmetry in $y$, i.e., BICs 3-5 in Table~I and BICs
4-6 in Table~II,  are robust. Those BICs with $\beta_0 =0$ in structures
without a reflection symmetry in $y$, i.e., BIC1 and BIC2 in
Table~III, are nonrobust (ordinary ones), but can still be preserved by adjusting a
second parameter when the first parameter is varied. Those FP-BICs with
$\beta_0=0$ in structures that are symmetric in $y$, i.e., BIC1 and
BIC2 in both Table~I and II, are also nonrobust
but they are 
the so-called super-BICs. A super-BIC may bifurcate to a par of normal robust
BICs with nonzero and opposite $\beta_0$ when a parameter is
varied. The BIC3 in Table~II can be regarded as the robust BICs
bifurcated from BIC1 or BIC2, also in Table~II, as one parameter is
varied. However, it should be emphasized that the existing theories on
robustness, parametric dependence and super-BICs, do not imply the
existence or nonexistence  of FP-BICs near total reflections.
Our work is a numerical study. Further
theoretical studies are needed to clarify the connection between total
reflections and FP-BICs. 

\section*{Acknowledgments}
The authors acknowledge support from the Research Grants Council of Hong
Kong Special Administrative Region, China (Grant No. CityU 11307720).

\end{document}